# On the role of Calcium in inducing superconductivity – the study of La-2125 superconductors

S. Rayaprol[+] and D. G. Kuberkar

*Department of Physics*

*Saurashtra University*

*Rajkot – 360 005*

INDIA

[+]***Corresponding Author***

*Email:*     sudhindra@rayaprol.com / rayaprol@gmail.com

**ABSTRACT**

Ever since high $T_c$ superconductivity was discovered in La-based mixed oxide system by Bednorz and Muller, enormous efforts have been put in by several researchers around the world in understanding the origin and mechanism of superconductivity in these, as well as in systems derived from them. It is a proven fact that the superconductivity in RE-123 superconductors is governed by the oxygen content, which in turn is responsible for the carrier concentration in the system. Due to their dependence on oxygen content, RE-123 superconductors undergo structural transformation from orthorhombic to tetragonal as a function of oxygen content making them very difficult compounds to work with, in terms of technological applications, such as device fabrication. Hence, it would be interesting to obtain a stable compound whose superconducting properties, are not only insensitive directly to oxygen content but, having dependence of its carrier concentration and $T_c$ on the nature and amount of the substituted cation. In the present work, we focus our investigations for such a compound, which we derive from a tetragonal RE-123 superconducting system. In this chapter, we present a brief review of the studies carried out on La-2125 compounds to elucidate the role of dopants in modifying the superconducting properties and establish a structure – property correlation in them.

1. **<u>INTRODUCTION</u>**

The discovery of superconductivity by Kammerling Onnes [1] paved way for condensed matter physicists and material scientists in synthesizing new compounds and discovering this novel phenomenon in a variety of compounds such as pure metals, alloys and mixed oxide systems. The discovery of high temperature superconductivity in a mixed oxide La-Ba-Cu-O compound by Bednorz and Muller in 1987 [2, 3] lead to an explosion in research on these novel materials, with several groups around the world began investigating the effects of elemental substitutions and different processing conditions on the structure and superconducting properties of this oxide. As a follow up of this effort was the discovery of superconductivity above 90 K in Y-Ba-Cu-O system by Paul Chu and co-workers [4]. These results were reproduced quickly and it was observed that the 90 K phase was $YBa_2Cu_3O_7$ leading to tremendous research activities in this field all over the world. The cycle of experimentation with elemental substitution and different processing conditions repeated after this, leading to the discovery of superconductivity above 100 K in Bi-Ca-Sr-Cu-O by Maeda et al [5] and Tl-Ca-Ba-Cu-O system Herman et al [6, 7].

Although the initial hype in the flurry of research activities in these high temperature superconducting systems has died down now, there is still a considerable interest in the study of these fascinating oxides, which has been largely attributed to the tunability of properties by elemental substitution for increasing superconducting transition temperature ($T_c$) or critical current density ($J_c$) or critical field ($H_c$). Several theories have been put forward to explain the origin of superconductivity and to understand the physical characteristics of these oxides [8], yet the complete picture is far from clear.

Since the discovery of high temperature superconductors (HTSC) several thousand papers might have appeared in various journals and books, giving a very comprehensive picture of various studies carried out and new results are being added to this mega list on a regular basis. In context with scope of this chapter, we restrict our focus on the appearance of superconductivity in Y-123 type triple perovskites, which can be built in different ways [see for example 9 - 13].

Superconductivity can be induced in a non-superconducting, anti-ferromagnetic $La_2Ba_2Cu_4O_z$ (La-224) by adding equal amounts of CaO and CuO. The resultant stoichiometric composition $La_2CaBa_2Cu_5O_z$ (La-2125) shows a maximum $T_c \sim 78K$ [14]. The structural investigations on polycrystalline La-2125 system carried out using X-ray diffraction (XRD) and neutron diffraction (ND) techniques shows cationic intermixing at various crystallographic sites, which are responsible for superconductivity. The superconducting transition temperatures ($T_c$), measured using resistivity ($\rho$) and magnetization measurements ($\chi$), show a strong dependence of $T_c$ on the density of mobile charge carries, i.e., "hole concentration" in conducting $CuO_2$ sheets ($p_{sh}$). The Ca-doping at $La^{3+}$ site plays an important role in "turning on" of superconductivity into this system [15-17]. The present chapter presents the studies on rare earth doping at La site in $La_{2-x}RE_xCa_{2x}Ba_2Cu_{4+2x}O_z$ [La-2125 for RE = Pr, Gd, Y, Nd etc] system along with the simultaneous increase in Ca content, in order to understand the structural, transport and magnetic properties of systems studied. A comparative study between isostructural RE-123 and La-2125 system has been carried out to understand the structural aspects of these triple perovskites. In the following sections of this chapter we discuss the structural, magnetic and transport properties of these interesting compounds.

## 2. GENERATION OF Y-Ba-Cu-O (1-2-3) STRUCTURE

In order to understand the La-2125 structure, it is important to understand the structure of the parent compound first. Soon after the discovery of superconductivity in Y-Ba-Cu-O, it was found that the phase responsible for $T_c > 90$ K has a cation composition of Y: Ba: Cu in the ratio of 1:2:3 and hence these compounds are now called the "***1-2-3 superconductors***".

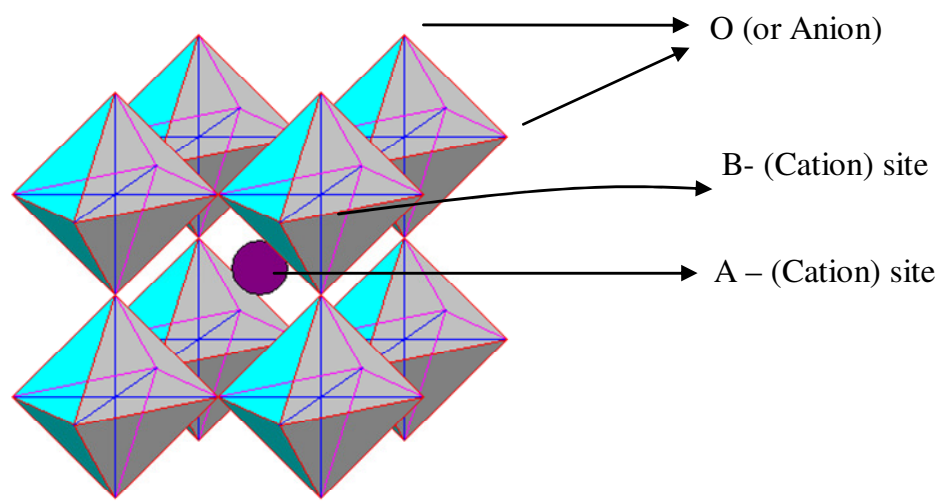

**Fig. 1**      **Simple cubic perovskite structure**

The unit cell dimensions determined by X-ray and neutron diffraction measurements revealed a structure related to cubic perovskite with one of the cube axes (c-axis) tripled [18] hence termed as "*triple perovskites*". The cubic perovskite structure, $ABO_3$ (see Fig. 1), is identified by two cationic sites, one at the centre of a cage formed by the corner-sharing oxygen octahedra (A-site) and the other lies at the center of the oxygen octahedra (B-site). The A-site can accommodate larger cations, where as B-site can accommodate smaller cations. Since the A-site can accommodate larger ions, the $Y^{3+}$ or other rare earth (usually $RE^{3+}$) or $Ba^{2+}$ ions can be placed at these sites, whereas the smaller $Cu^{1+}$ or $Cu^{2+}$ ions are accommodated in the B-site. A triple perovskite can be achieved by stacking three perovskite blocks with the central block having Y or RE ion at the A-site, where as the top and bottom blocks will have a Ba ion at the A-site. This arrangement approves the stoichiometric composition of 1Y:2Ba:3Cu, and can be seen clearly in the structure of the oxygen deficient triple perovskite, as shown in the figure 2.

## 2.1  *The Orthorhombic Y-Ba-Cu-O structure*

The three oxygen ions per unit cell in $ABO_3$ perovskite block leads to nine oxygen ions per unit cell in a triple perovskite structure. Formal balancing of the charges on cations considering $Cu^{3+}$ leads to 8 oxygen per unit cell, whereas if we assume $Cu^{2+}$ or $Cu^{1+}$ then 6.5 and 5 oxygen ions are required per unit cell respectively, which is less than the observed oxygen content per unit cell.  Hence, the triple perovskite structure of the 1-2-3 superconductors is also called the "oxygen deficient perovskite" relative to the ideal perovskite structure. Analysis of the X-ray diffraction from small single crystals extracted from polycrystalline samples confirmed and refined the basic positions of the cations and identified where the oxygen deficiency was accommodated in the structure [19 - 22]. The X-ray data clearly showed that the Y (or rare earth) ion is surrounded by eight 'O' ions only, instead of twelve in the case of an ideal perovskite structure. Oxygen deficiency was also noted in the basal copper plane between the Ba ions (see figure 2). The oxygen sites in this plane, which lies at the cell edges, are only half-filled [20 – 22]. Initially there were few discrepancies in the structures reported for the 1-2-3 compounds by studies on the polycrystalline samples, [20-21, and 23-26] and single crystal studies [19 - 22]. In a very rare occurrence, the structure of 1-2-3 type high temperature superconductors was first correctly determined by the Rietveld analysis of neutron powder diffraction data [27]. The advantage of using Rietveld refinement technique [27, 28] is that the structure can be elucidated directly from powder diffraction data (Neutrons or X-rays) by fitting the data with a simulated diffracted pattern assuming a near perfect starting model. Using neutron diffraction for structural elucidation is more rewarding since the larger scattering cross-section for oxygen by neutrons relative to x-rays can be advantageous as it reduces the need for having large single-crystals. The Rietveld analysis in several studies [20, 21, 23-26] have consistently shown that the structure of Y-Ba-Cu-O is **orthorhombic** with a unit cell in which *b* is slightly greater than *a* and *c* is roughly *3a*. This structure was assigned to the space group **Pmmm, ($D_{2h}^1$)** with one formula unit per unit cell and the representative lattice parameters, *a* = 3.827 Å, *b* = 3.882 Å and *c* = 11.682 Å. The effect of oxygen ordering in the basal plane of the orthorhombic Y-Ba-Cu-O structure is to occupy one of the O sites along a cell edge and leave the other site vacant, which corresponds to the presence of –Cu-O-Cu-O chains along the *b* direction. The vacancy on the O-site in the cell edge, causes compression of unit cell along *a* and elongation of *b* due to the filled cell edge, producing an orthorhombic unit cell. This ordering puts the copper ions in the basal plane of the structure, i.e. at the centre of square arrangements of the oxygen ions. The square planar arrangements of copper and oxygen ions are then linked together by sharing their corners to form "chains" along the *b* axis of the structure. The effect of vacant oxygen sites around the Y ion can be seen in Figure 2.

      Copper ions are in five-fold coordinated square pyramidal sites in the Cu-O plane because of vacant oxygen sites around Y. The Cu-O bonds in the planes are plucked throughout the *a-b* plane due to the linking of square bases of the pyramids at their corners. The chains and planes in the structure are linked through the oxygen ions that lie at the apices of the square pyramids. The Y ion, which lies at the center of the cell, is coordinated by 8 oxygen ions that forms slightly distorted square prism. The Ba ion is in 10-fold coordination and is shifted slightly towards the Y ion relative to its position in the ideal perovskite structure.

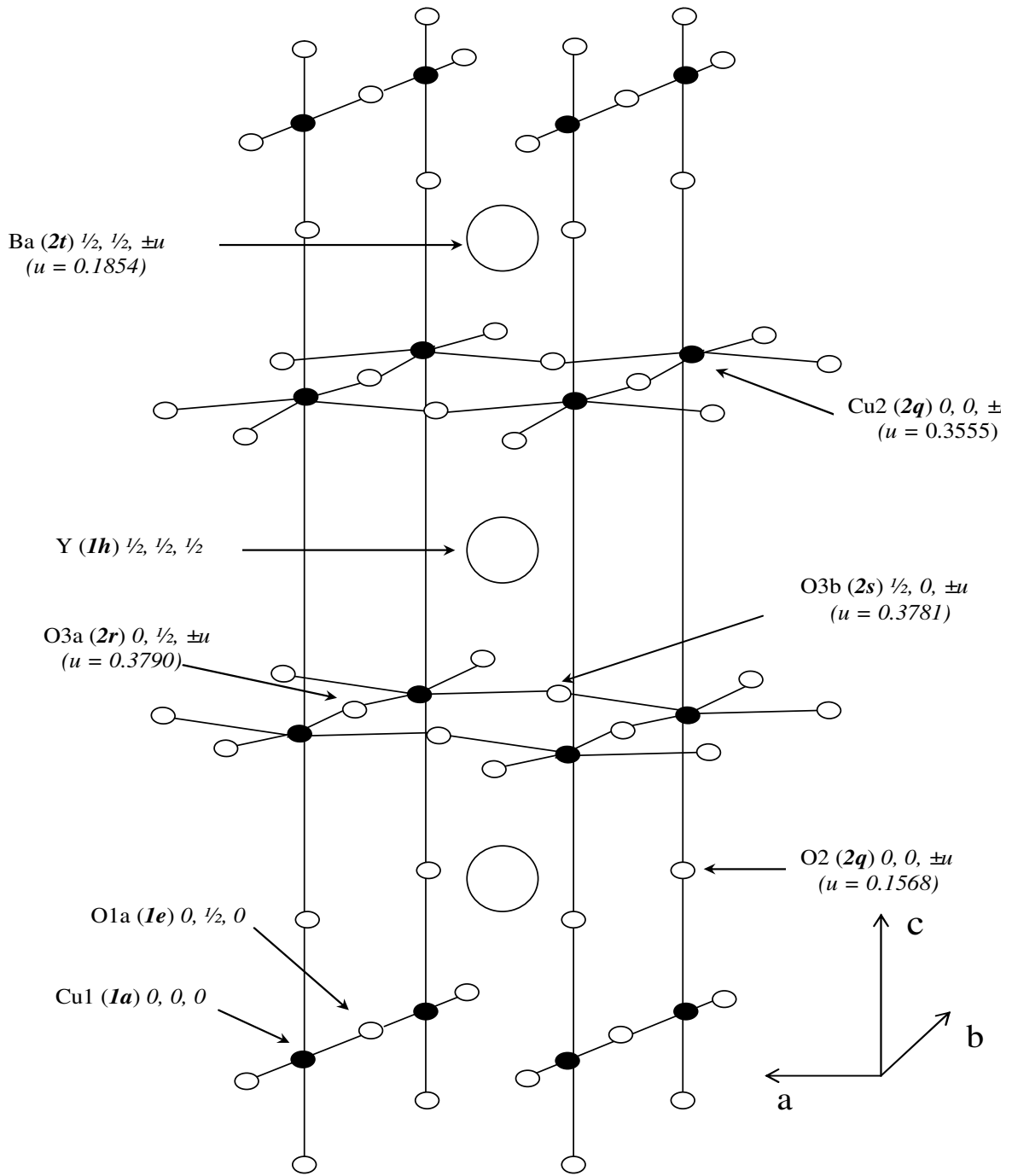

**Fig. 2** Structure of the orthorhombic unit cell of $YBa_2Cu_3O_{7-\delta}$ *(Pmmm)*. Wyckoff notations and atomic positions are shown in parenthesis.

## 2.2 *The Tetragonal Y-Ba-Cu-O structure*

The oxygen content in the Y-Ba-Cu-O structure depends upon the processing conditions also. The tetragonal phase is stable in these compounds above 700°C and on slow cooling from 700°C the tetragonal to orthorhombic transition takes place. However, the tetragonal phase is more persistent if the sample is quenched from higher temperatures or cooled in a reducing atmosphere. Cationic intermixing of RE-Ba, Ca-RE, and Ca-Ba has also resulted in the tetragonal structure formation [29, 30].

All the essential features of a tetragonal Y-Ba-Cu-O structure are shown in the figure 3. This is a tetragonal version of the orthorhombic $YBa_2Cu_3O_7$ (superconductor) formed by removing the chain oxygen's so that none remain on the basal plane. Rietveld refinements on neutron diffraction data of oxygen deficient materials show that oxygen is lost primarily from the O1 site in the chains [31-35]. There is one formula unit per unit cell. Yttrium and one copper atom are in special positions, and the remaining atoms are all in general positions with a single undetermined parameter associated with *z* coordinate of each. This semi-conducting compound was assigned the space group **P4/mmm** ($D_{4h}^1$). The general atomic positions are also shown in Figure 3.

In these oxygen-reduced compounds, there are only six oxygen atoms per unit cell and the chain site is completely empty, which reduces the coordination of Cu1 atom to two. This copper coordination led to the suggestion that tetragonal 123 compounds contains $Cu^{2+}$ ions in the planes and $Cu^{1+}$ ions in the two-fold coordinated sites [34, 35]. The unusual linear two coordination of copper with two Cu1-O2 bonds along with the charge balance considerations, suggests that Cu1 site contains $Cu^{1+}$ ions.

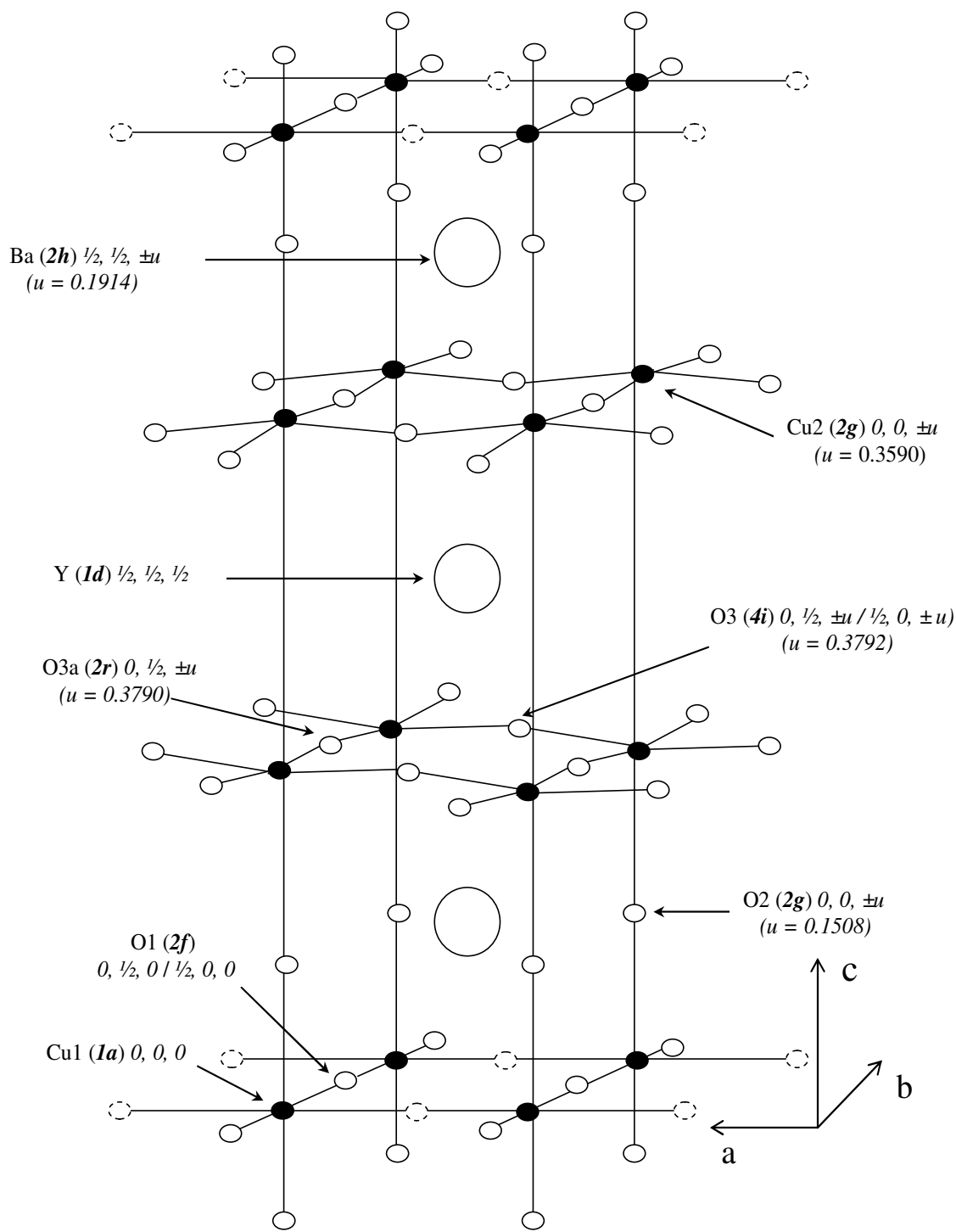

**Fig. 3**     Structure of tetragonal unit cell of $YBa_2Cu_3O_6$ (*P4/mmm*). Wyckoff notations and atomic positions are shown in parenthesis.

The Y-O and Ba-O distances are typical of those found in other oxides. Planes of Cu2 and O3 are the important features of this 1-2-3 structure. Each Cu2 has four nearest neighbors O3 oxygen's, as well as a fifth weakly bonded apical O2. Each O3 is displaced along *c* towards Y, so Cu2-O3 layers are buckled. Both O2 and O3 are in distorted octahedral coordination. The O2 coordination group includes one short Cu1-O2 bonds (~1.8 Å) and a long Cu2-O2 bond (~2.45Å). Thus, there is little coupling between Cu2-O layers and Cu1. Oxygen content in the tetragonal Y-Ba-Cu-O ($YBa_2Cu_3O_6$) may exceed six per unit cell, depending upon the temperature and annealing conditions. Excess oxygen may enter the O1 positions at (*0, ½, 0*), which can hold up to two oxygen per unit cell. The O1 position is thus typically about ¼ occupied in the Y-Ba-Cu-O phase at room temperature [36-40]. An important consequence of greater O1 occupancy is increased average valence of Cu1. Excess oxygen in a Y-123 structure ($z > 7.0$) also results in tetragonal structure [9, 18], classifying them in 'over-hole doped' superconductors. Most of the divalent cation doped systems are found in this region of 'hole concentration' only. But increasing oxygen content beyond 7.0 has often resulted in loss of superconductivity [9 and references there in].

## 3.  **SYNTHESIS, STRUCTURE & PROPERTIES OF La-2125 COMPOUNDS**

### *3.1  Introduction*

The compounds of the type $La_3Ba_3Cu_6O_{14+\delta}$ (La-336), $La_3CaBa_3Cu_7O_z$ (La-3137), $La_{1.5}Ba_{1.5}Cu_3O_z$ (La-112), $LaBaCaCu_3O_9$ (La-1113) etc have been found to be isostructural to the La-Ba-Cu-O tetragonal structure. Among these compounds, the Ca containing compounds have been found to be superconducting with $T_c \sim 80$ K [9, 41-46]. The interest in studying these tetragonal superconductors is due to the role played by the dopants, especially 'Ca', in introducing superconductivity without modifying the structure. The superconductivity is found to be dependent on the hole concentration and thus providing a tool for possible technological applications. In this chapter, we focus on the studies carried out on La-224 compounds, which show increase in superconductivity, when equal amounts of CaO and CuO are added to the La-224, to get a superconducting La-2125 phase.

### *3.2 Synthesis*

The polycrystalline samples of La-2125 compounds are synthesized by solid-state reaction method, employing the synthesis route generally followed for preparation of high temperature superconductors. The high purity starting compounds, such as $RE_2O_3$, $CaCO_3$, $BaCO_3$ and $CuO$ are initially homogenized by grinding under acetone. These compounds are then calcined at 800 °C for about one day, and then thoroughly re-grinded and palletized before sintering for about a day in air, in the temperature range of 920°C – 950°C, which completes the reaction process. For oxygen homogeneity, the sintered pellets are first annealed in flowing $N_2$ gas for about 6 hours at 920°C – 950°C, and then in dry $O_2$ for 6 -10 hours followed by slow cooling to room temperature. It is observed that starting compounds, particularly $RE_2O_3$ and carbonates if pre-heated gives good results.

### *3.3 Structure*

Earlier reports on the La-336 type compounds [46, 47] emphasized the role of intergrowths of multiple oxygen deficient perovskite layers and "rock-salt" type layers in the crystal structure of high temperature superconductors. Contrary to these reports, a single-phase triple perovskite structure was clearly seen in the XRD patterns. The size of the unit cell and the X-ray patterns are conclusive proof that the rock-salt type layers of $CaCuO_2$ is not added to the structure, but Ca and Cu are distributed among the crystallographic sites within the structure.

Structural analysis of La-2125 type compounds by XRD and ND measurements have conclusively shown the cationic intermixing at La and Ba sites, resulting in the addition of holes with increasing Ca content, *vide infra*.

Several efforts to prepare a pure (without partial RE substitution) $La_2Ca_1Ba_2Cu_5O_z$ did not yield repetitive results, and hence were not pursued in detail. The substitution of RE at La site, in small quantities (up to x = 0.5) have been studied, and a relationship between substituted RE and Ca content (RE = ½ Ca) has been maintained in all the systems studied.

The genesis of the structure of La-2125 type compounds is based on the stoichiometric arrangement of $La_2Ba_2Cu_4O_z$ in La-Ba-Cu-O (1-2-3) form and then adding CaO and CuO to it. The La-224 is written in La-123 form as:

$La_2Ba_2Cu_4O_z$ → $(La_2Ba_2Cu_4O_z) \times (3/4)$

→ $La_{1.5}Ba_{1.5}Cu_3O_{z'}$ (where z' = 3z/4)

→ $La_1(Ba_{1.5}La_{0.5})Cu_3O_{z'}$ → La-123         --- (1)

The Rietveld analysis of the powder neutron diffraction data of La-224 compound has confirmed the stoichiometric composition as shown in equation (1) and a structure similar to a tetragonal La-123 structure [16]. Similarly, the La-2125 compound can also be shown in La-123 composition as:

$La_2Ca_1Ba_2Cu_5O_z$ → $(La_2Ca_1Ba_2Cu_5O_z) \times (3/5)$

→ $La_{1.2}Ca_{0.6}Ba_{1.2}Cu_3O_{z'}$ (where z' = 3z/5)

→ $(La_{0.6}Ca_{0.4})(Ba_{1.2}Ca_{0.2}La_{0.6})Cu_3O_{z'}$         --- (2)

When $RE^{3+}$ (where RE is rare earth, such as Dy, Nd, Pr, Gd, Er, Y etc) is substituted at $La^{3+}$ site, then, for x = 0.5, the La-2125 compound with $RE^{3+}$ can be represented as:

→ $(La_{0.3}RE_{0.3}Ca_{0.4})(Ba_{1.2}Ca_{0.2}La_{0.6})Cu_3O_{z'}$         --- (3)

in normalized (to La-123) form. Independent neutron diffraction measurements on RE = Dy and Nd, have confirmed the stoichiometric composition of $La_{1.5}Dy_{0.5}Ca_1Ba_2Cu_5O_z$ and $La_{1.5}Nd_{0.5}Ca_1Ba_2Cu_5O_z$ can be represented as shown in equation (3) for performing structural analysis. Compounds with x > 0.5, in the said stoichiometry, were, found to crystallize in another isostructural phase La-1113. The simultaneous displacement of Ca on both RE and Ba sites is an interesting feature of these compounds [14-17]. Factors (3/4) and (3/5) used in equations 1, 2 & 3, represent the ratio of number of copper ions in RE-123 to the number of copper ions in La-224 and La-2125 unit formula respectively. Equations 1, 2 and 3 shows the normalized La-224, La-2125 and RE substituted La-2125 phases respectively, which help in understanding the distribution of various cations in the normalized form. Figure 4 shows a typical neutron diffraction pattern for the La-2125 compound recorded at room temperature. The observed data has been fitted into a tetragonal triple perovskite structure with P4/mmm space group. The red colored continuous line shows the calculated pattern. The blue line indicates the difference between the observed and calculated pattern. Bragg peak positions are indicated by tick marks.

In the La-2125 normalized form, given by equation (2), $Ca^{2+}$ is distributed at both $La^{3+}$ and $Ba^{2+}$ sites along with the concomitant displacement of $La^{3+}$ onto $Ba^{2+}$ site. As discussed earlier, in the RE-123 unit cell we find two types of cations, 'A' (e.g., $Y^{3+}$ or trivalent lanthanide cations, which have typical ionic radii ~ 0.9 – 1.3 Å for coordination number (CN) 6) and 'B' type ($Cu^{2+}$ with ionic radii ~ 0.73. Owing to the similar ionic radii of $La^{3+}$ and $Ca^{2+}$ (1.03Å and 1.00 Å, CN 6 respectively), $Ca^{2+}$ is expected to prefer $La^{3+}$ than to $Ba^{2+}$ site (~ 1.35 Å CN 6) [50].

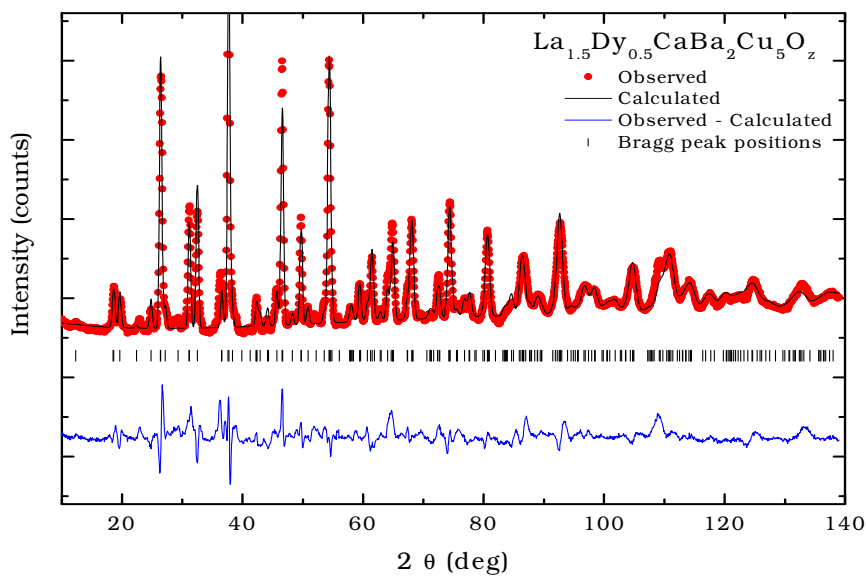

**Fig. 4 Neutron diffraction pattern for $La_{1.5}Dy_{0.5}CaBa_2Cu_5O_z$ recorded at room temperature and fitted to a tetragonal triple perovskite unit cell (La-123 type) using Rietveld method.**

Interestingly, the structural analysis of the La-2125 type oxides confirms the presence of Ca at Ba site also. Though $Ca^{2+}$ at $La^{3+}$ is 'hole doping', $La^{3+}$ at $Ba^{2+}$ is 'hole filling' it has been found that the simultaneous substitution of Ca at La and La at Ba is non-compensatory which can be seen by the increase in hole concentration and $T_c$ [15, 53]. Rietveld analysis of X-ray and neutron diffraction data of these compounds have clearly established the tetragonal structure, isostructural to the tetragonal La-123 compound with ***P4/mmm*** ($D_{4h}^1$) space group. Figure 5 show a typical La-2125 unit cell with the atomic positions of dopants which are also found to be occupying the rare-earth, and Barium sites [14-17].

Figure 6 shows the variation of unit cell parameters and volume with increasing dopant concentration. The unit cell parameters decreases with increasing smaller ion dopant concentration, leading to decrease in lattice parameters and subsequently leading to the contraction of the unit cell volume, in accordance to the Vegard's law. This is a clear indication of the fact that smaller ions (such as $Ca^{2+}$ and $RE^{3+}$) are replacing bigger ions in the lattice, resulting in the shrinking of the unit cell dimensions. It is interesting to note that superconductivity in these compounds increases with increasing dopant concentration. In the present context, chemical pressure created by the doping of smaller ions, helps in inducing and enhancing superconductivity as the unit cell volume decreases. Since the rate of decrease of the unit cell dimensions is same for different $RE^{3+}$ ions, the plot of c/a vs. x remains same for all RE, we believe that increasing dopant concentration (particularly, Ca) plays an important role in inducing superconductivity in the non-superconducting parent compound.

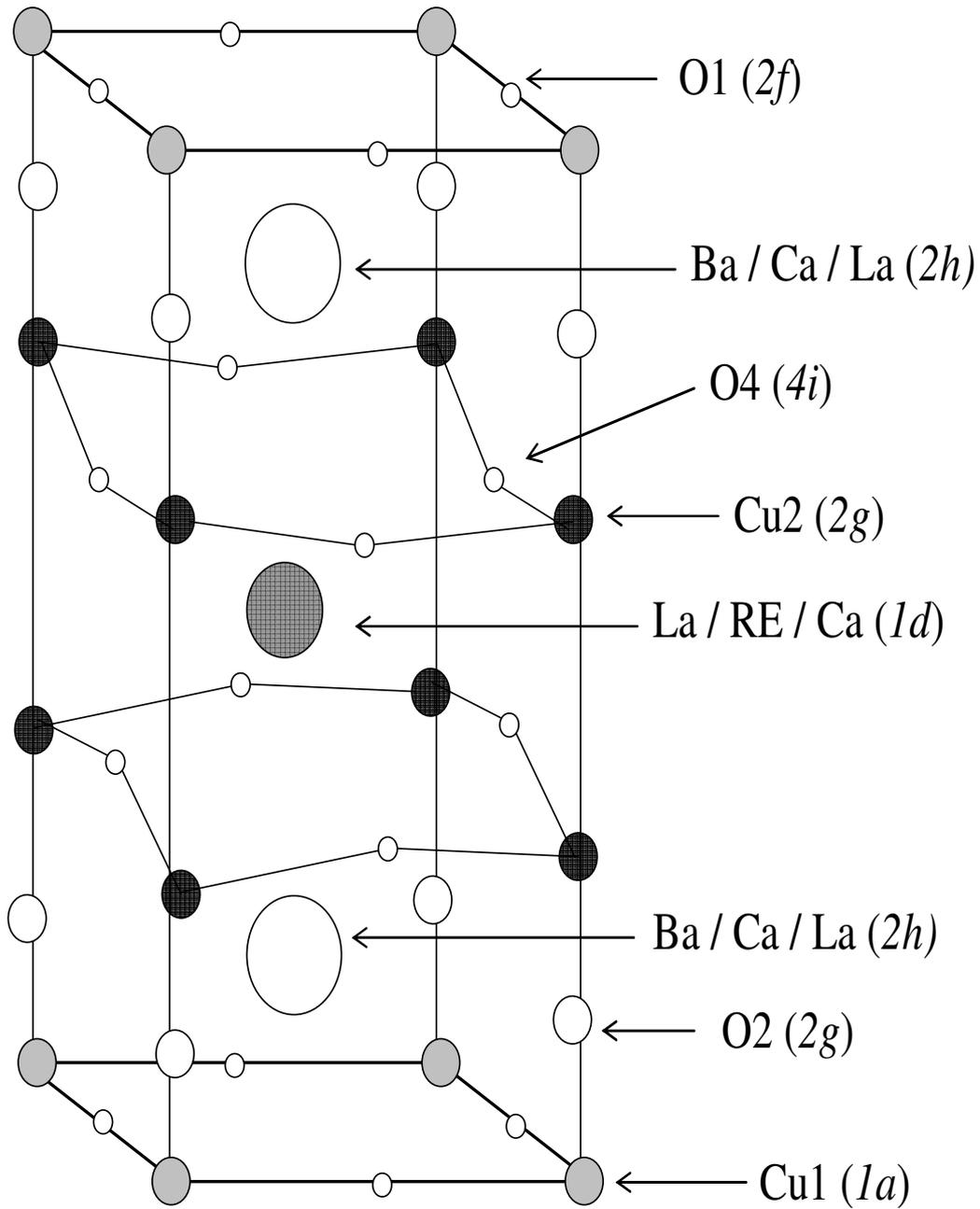

**Fig. 5** **The structure of the La-2125 unit cell, derived from the tetragonal La-123 structure. Wyckoff positions are represented in parenthesis.**

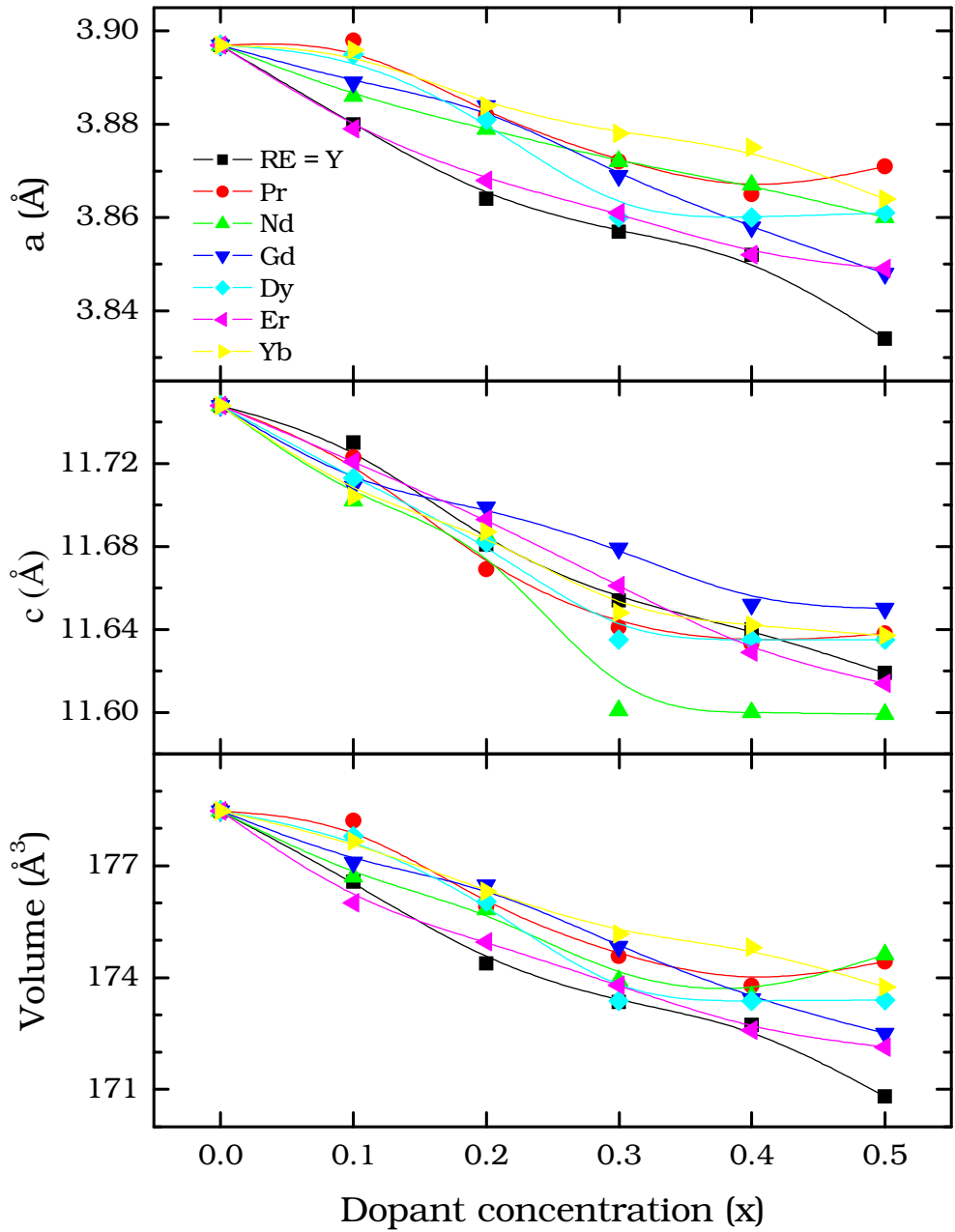

**Fig. 6** **Decrease in the lattice parameters (*a*, *c* and *Volume*) with increasing dopant concentration for different RE substituted La-2125 compounds.**

## 3.3 Bond lengths, transition temperature, oxygen content and hole concentration:

The effect of Ca doping on bond-lengths in the superconducting Cu-O block around La-atom, for various RE substituted La-2125 systems can be seen in Figure 7.

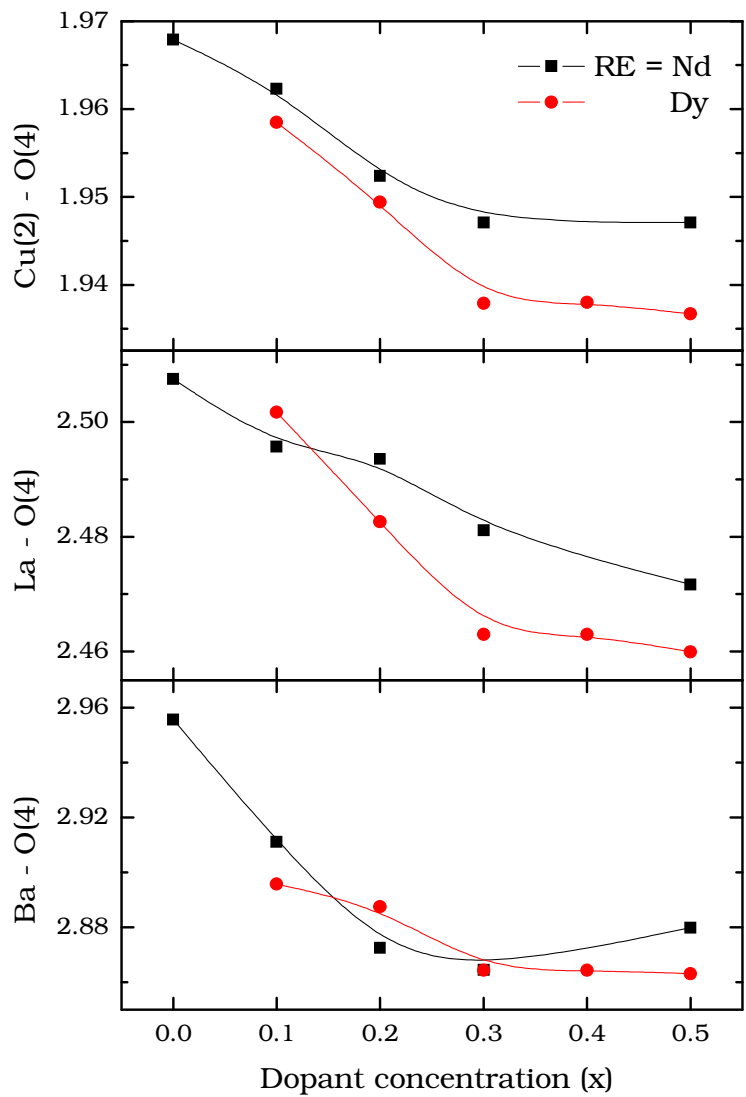

**Fig. 7** **Variation in selected bond lengths with increasing dopant concentration.**

The gradual decrease in the bonds between the two Cu-O planes sandwiching the La atom with increasing RE and Ca concentration, accompanied by subsequent increase in $T_c$ is indicative of the strong structure-property correlations in these compounds.

The superconducting transition temperature, $T_c$ is usually determined by the resistivity and susceptibility measurements. The similarity of both measurements shows the superiority of the sample quality. Precise values of superconducting transition temperatures can be gauged by the first derivative of their respective susceptibility curves [See for example Ref. 52 and references therein]. In Figure 8, we show the $T_c$ (K) measured by electrical (dc resistivity) and magnetic (ac /dc susceptibility) methods obtained for different samples, as a function of x. The difference in $T_c$ (K) obtained by these two independent methods is ±1 K. The maximum $T_c$ between 75 to 78 K is observed for x = 0.5, i.e., the La-2125 phase for all RE except Pr. The maximum $T_c$ observed in the case of Pr is ~58 K for x = 0.5 . It is believed that the hybridization and electron localization as a consequence of interaction between the Pr moment (3+ or 4+) and Cu-O plane is responsible for such an effect. It is interesting to observe here that in addition to this phenomenon, hole doping in form of $Ca^{2+}$, takes places, resulting in the increase in $T_c$. Though the rate of increase of $T_c$ for RE = Pr is similar to other RE, the maximum $T_c$ achieved is not same (See Fig 9) [53-54].

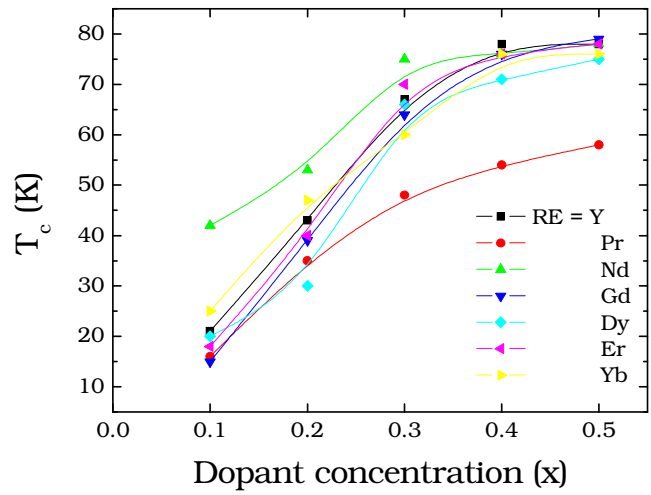

**Fig. 8**      $T_c$ vs. $x$ for $La_{2-x}RE_xCa_{2x}Ba_2Cu_{4+2x}O_z$.

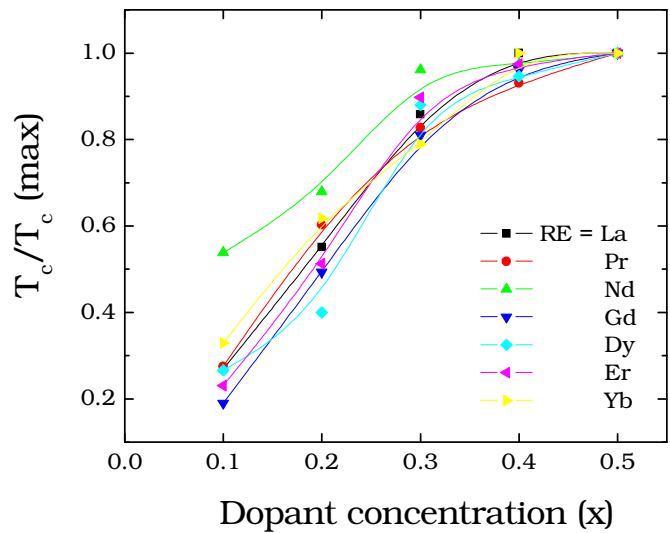

**Fig. 9**      **Transition temperature normalized to maximum $T_c$. The lines passing through the data points are guides for eyes only.**

We have seen the importance of oxygen in structural stability of the RE-123 superconductors. A superconducting orthorhombic structure has an ideal value of $z \sim 6.9$. Oxygen content above 7.0 drives the system into 'over-hole doped' regime thus suppressing $T_c$ from ~90 K to ~80 K. In this region of oxygen content value, $T_c$ is almost independent of absolute value of $z$, it rather depends more on $p_{sh}$ [13, 15]. In the presently studied systems, the structural stability has been seen for all the compounds prepared and reported. Several structural studies have proved that there is no structural transition [14, 16-17], or modification, even at low temperatures, below $T_c$ [55]. Hence, the role played by $Ca^{2+}$ substitution in affecting hole concentration and $T_c$, in La-2125 superconductors is very crucial.

In Figure 10, we show the oxygen content per formula unit, obtained from Iodometric double titration method [a detailed explanation is given in Ref. 52]. The charge carrier concentration (holes) per formula unit was derived from oxygen content values. Increase in oxygen content with increasing dopant concentration results in increase in hole concentration in the unit cell, resulting in the increase of mobile charge carriers in the conducting Cu-O sheets (Fig 10).

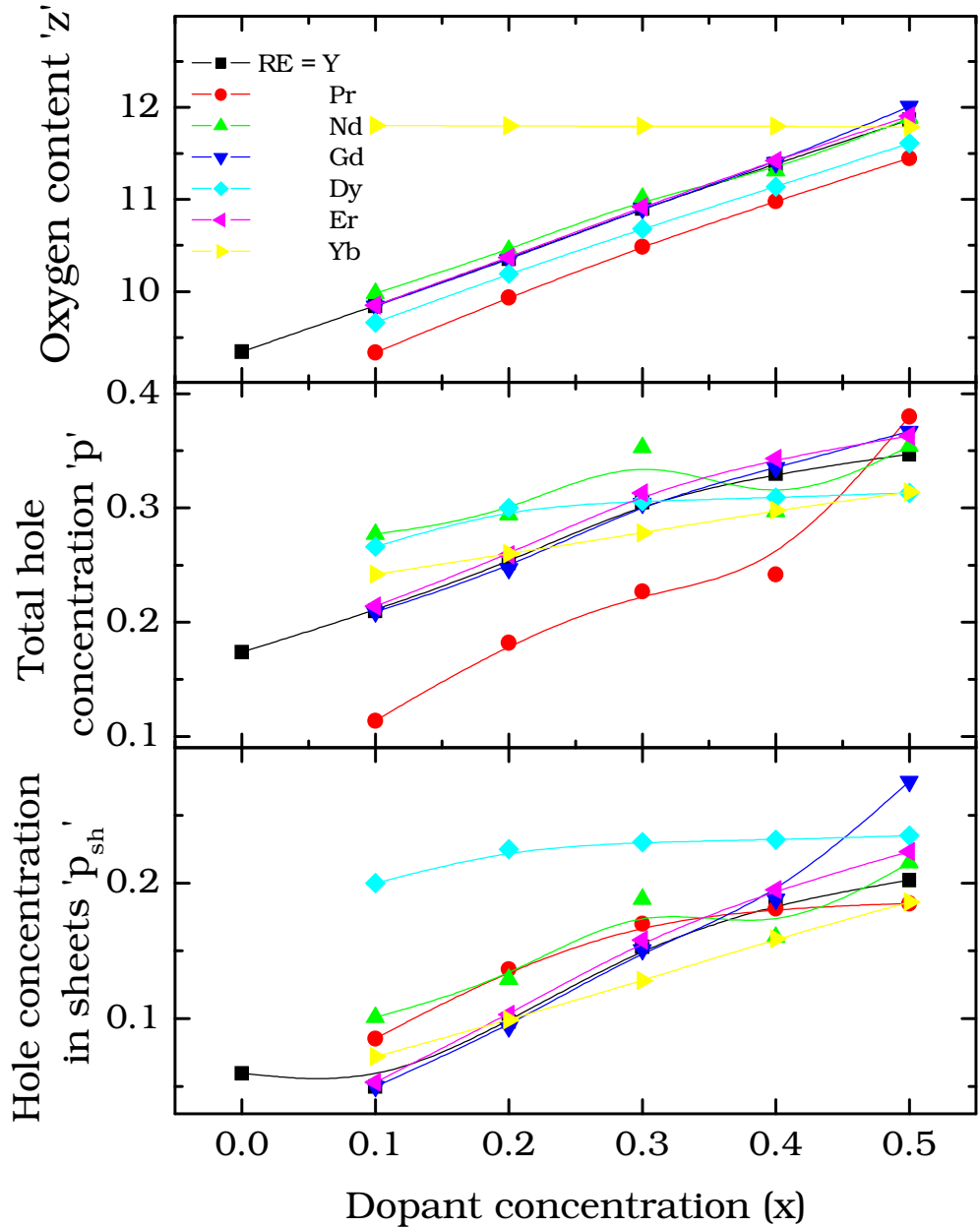

**Fig. 10** Oxygen content per formula unit obtained from Iodometric double titration experiment. Hole concentration has been calculated from oxygen content values.

## 4. Concluding remarks

A combined look at the structural (unit cell volume Fig. 6), superconducting transition temperature (Fig. 8) and oxygen content values (Fig. 10) clearly establishes the importance of $Ca^{2+}$ in inducing superconductivity in a non-superconducting parent compound. Our studies, involving different $RE^{3+}$ ions clearly establishes that there is no effect of $RE^{3+}$ moment (except for partial suppression of $T_c$ in case of Pr) on superconducting behavior. The onset of superconductivity in all the samples discussed above can be consistently understood by the hole doping, which is responsible for the bridging of the two Cu-O sheets, around the central RE ion site, in the central perovskite block of the triple perovskite structure. The bridging of Cu-O planes in tetragonal superconductors was given by Gu et.al [56], and explained for La-2125 superconductors by Rayaprol et.al [17].

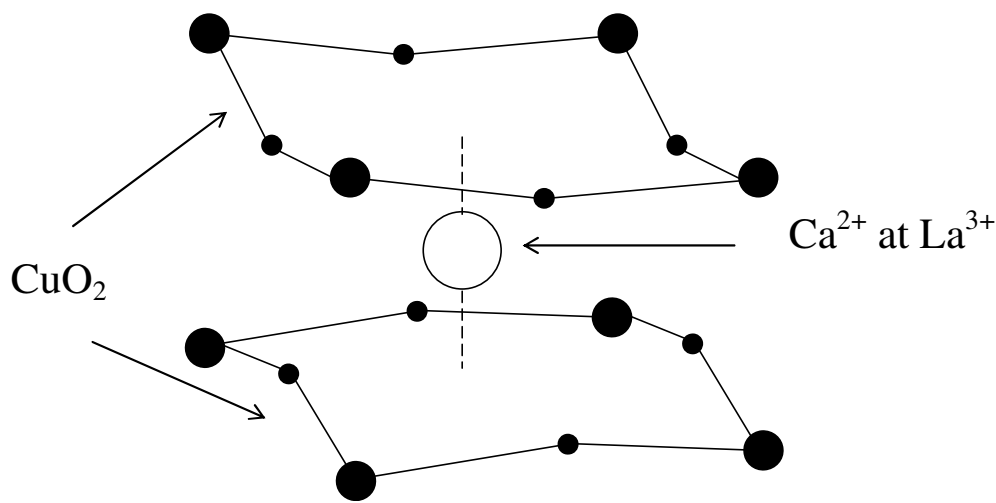

**Fig 11**     **The superconducting perovskite block**

Figure 11 shows the conducting block in the tetragonal perovskite structure. Both the Cu-O sheets are dimpled in the high Tc superconductors [57]. The dimpling of the Cu-O sheets is further enhanced by the substitution of smaller ions, such as Ca for La (or RE for La) resulting in the shrinking of the a-b plane, leading to the overall decrease in the cell volume. The shrinking of a-b plane results in bridging of the $CuO_2$ sheets (figure 11) resulting in a shorter path for charge carriers to conduct through. Again looking at Fig 7 and 8 in this context, it clearly establishes that increasing dopant concentration results in more bridging of the Cu-O planes in La-2125 compounds resulting in increase in superconducting transition temperatures. The buckling or dimpling of the Cu-O sheets has pronounced effects on the flux pinning in these compounds which increases with increasing dopant concentration [17, 54].

Finally, a series of compounds in the stoichiometric equation, $La_{2-x}RE_xCa_{2x}Ba_2Cu_{4+2x}O_z$ has been synthesized from $La_2Ba_2Cu_4O_z$, by adding equal amounts of CaO and CuO, along with different RE substituions at La site. No structural transition takes place in entire doping range (up to x = 0.5), but $T_c$ increases with increasing dopant concentration, establishing a relationship between hole concentration and $T_c$. No effect of RE moment on $T_c$ was found except for the case of Pr, where a mixed valence Pr may be responsible for suppression of $T_c$. The present study clearly establishes that $Ca^{2+}$ acts as a hole dopant in the La-2125 tetragonal compounds, driving them from non-superconducting (for x = 0.0) to a superconducting (x = 0.5) system with a maximum $T_c$ ~ 78 K. The defects created by the substitution of smaller ions for bigger ions, acts as flux pinning centers resulting in increase in critical current densities[17, 54].

We have presented here a preliminary study on these interesting compounds. Efforts have been made to synthesize thin films on these compounds using Pulsed laser deposition technique and study the effects of ion beam irradiation in modifying the superconducting character of these compounds [58]. Low temperature neutron diffraction experiments have also been performed. The role of copper valence has been studied in detail by bond valence sum calculations, results of which will be published separately.

**5.    Acknowledgements**

Authors would like to express gratitude to UGC-DAEF (formerly IUC-DAEF), NSC (New Delhi) and other funding agencies for financial assistance in carrying out the experimental work and stipendiums. We also thank our collaborators Dr. M. Ramanadham, Keka Chakraborty and P. S. R. Krishna from Solid State Physics Division at BARC (India) for neutron diffraction experiments and analysis.  We thank all former students and colleagues at Saurashtra University for discussing their work and providing results for this chapter. SR is indebted to Alexander von Humboldt Foundation, Germany for research fellowship.